\documentclass[useAMS,usenatbib,onecolumn,referee]{mn2e}
\usepackage{amssymb}
\usepackage{graphicx}
\usepackage{natbib}
\usepackage{bm}

\title[Compton scattering in a corona in two conditions]{Compton scattering in the optically thick uniform spherical corona around the neutron star in an X-ray binary in two conditions}
\author[Shi]{ ChangSheng Shi$^{1}$\thanks{E-mail:
shics@hainanu.edu.cn}\\
$^{1}$School of Science, Hainan University, Hainan 570228, China\\
}

\begin{document}

\date{Accepted ??; Received ??; in original form
 ??}

\pagerange{\pageref{firstpage}--\pageref{lastpage}} \pubyear{2017}

\maketitle

\label{firstpage}

\begin{abstract}
We consider the Compton scattering in the optically thick uniform spherical corona around a neutron star in an X-ray binary. In the scattering, the low energy seed photons ($0.1\ \sim\ 2.5\ {\rm keV}$) are scattered in low energy electrons ($2.5\ \sim\ 10\ {\rm keV}$) in the corona in two conditions, i.e. initial seed photons are scattered  in a whole corona and scattered in every layer of the corona that are supposed to be divided into many layers.
When the same number of input seed photons, the same corona parameters and the same energy distribution of all photons in the two conditions are considered, the approximately same number of output photons can be obtained, which means that there is approximately a transform invariance of layering the Comptonized corona. Thus
the scattering in the layers of a multi-layered corona is approximately equal to the scattering in the whole corona by dividing the whole corona into several layers.
It means that Compton scattering for the initial seed photons scattered in a whole optically thick spherical corona with uniformly distributed electrons also can be considered as that the multiple Compton scatterings take place in the layers of a multi-layered corona in order approximately, which can be used to explore some physical process in one part of a corona.
\end{abstract}


\section{Introduction}
\label{intro}
The photons diffusion in the plasma full of electrons was widely discussed, especially the process for the low energy seed photons scattered in low energy electrons. The process can be described by Kompaneets equations (e.g. Kompaneets 1957 ) and is described as Comptionization. Comptonization as a key physcial phenomenon often emerges in many literatures (Sunyaev \& Titarchuk 1979, Psaltis \& Lamb 1997, Lee \& Miller 1998). It can be used to explore the physics of some astrophysical phenomena in X-ray binaries (XBs) such as X-ray burst (e.g. Lapidus et al. 1986, Lewin 1993), quasi-periodic oscillations (e.g. Kumar \& Misra 2014, Karpouzas et al. 2020).

In an X-ray binaries, the neutron star (NS) can accrete matters from a companion star by two methods (disc accretion mode and wind accretion mode, Ghosh \& Lamb 1979, Shi, Zhang \& Li 2015, 2018). Then a corona in an XB may be formed (Galeev et al. 1979, White \& Holt 1982) and a lot of phenomena (e.g. X-ray burst, X-ray flare) can emerge in the corona. However, some physical processes can be produced in the whole corona and some (e.g. the movement of blobs, resonance) may only take place in a part of the corona (Wang et al. 1998). An oscillation accompanied with Compton scattering in a whole corona in an XB was discussed by Kumar \& Misra (2014), but the relation between  the Compton scattering in a whole corona and the scattering in one part of the corona is an open question.

In order to consider these physical processes that happen in a part of the corona, we calculate the photons number density in every layer of a multiple-layered corona and in the whole corona that are supposed to be divided into multiple layers. In the second section, we obtain that there is a transform invariance of layering an uniform spherical Comptonized corona if the number of input seed photons, the corona parameters, the energy distribution of all photons and the number of output photons are identical in the two conditions.
In section 3, we give an example on the Compton up-scattering for the low energy seed photons ($0.1\ \sim\ 2.5\ {\rm keV}$) scattered in low energy electrons $(2.5\ \sim\ 10\ {\rm keV})$ in an optically thick uniform spherical corona (with the optical depth: $1\ \sim\ 20$) around the NS in an XB. Then we obtain the conclusion that Compton scattering in a whole corona can be considered as the multiple Compton scatterings, i.e. Compton scatterings in the corona in the two conditions are approximately equivalent to each other. In the last section, the discussion and conclusions are presented.

\section{The transform invariance for a Comptonized uniform spherical corona divided into many layers}
Since some physical processes (e.g. oscillations) may be produced in one layer of a corona, it should be known how to describe the local state in every layer of a corona. In the radiation physics, some photons will be scattered if they travel across the medium with the optical depth ($\tau$). Then the intensity of the photons escaping from the medium can be expressed as  $I=I_0e^{-\tau}$, where $I_0$ is the intensity for the photons before the scattering. If seed photons are considered to be transported across n layers of the medium in proper order, the last intensity after the n scatterings is  $I=I_0e^{-\tau_1}e^{-\tau_2}...e^{-\tau_i}...e^{-\tau_n}=I_0 e^{-\tau}$, where $\tau_1,\ \tau_2,\ ...,\ \tau_i,..., \tau_n$ are the local optical depths and the subscripts $1,\ 2,\ ...i,\ ...,\ n$  express the 1st, 2nd, ..., ith..., nth layer respectively. In summary, the whole medium can be divided into many layers and the radiation process in the whole medium can be considered as multiple scatterings in the supposed multiple media in order. It can be obtained that equivalent result can also be approximately applied to the Compton scattering for an optically thick uniform spherical corona around a NS in an XB, which will be discussed below.

The exploration in this section bases on the assumption, i.e. that the same input initial seed photons scattered in the same electrons system should produce the same output photons for Compton scattering whenever the corona is or is not divided into layers according to the causality. In order to test the assumption, we explore two kinds of radiation processes of Compton scattering, i.e. seed photons are scattered in every layer of a multi-layers corona in order and are scattered in the whole corona. The radiation process in the whole corona has been discussed widely (Kumar \& Misra 2014, Karpouzas et al. 2019) but the first imaginary process was reported little.

Suppose that an optically thick spherical Comptonization corona with the depth $L$,  uniform electronic density $n_{\rm e}$, electronic temperature $T_{\rm e}$ around the neutron star in an XB can be divided into n layers with the depths $L_1,\ L_2,\ ...,\ L_i,\ ...,\ L_n$ and the volumes $V_1,\ V_2,\ ...,\ V_i,\ ...,\ V_n$ respectively. Thus the whole optical depth is the sum of local optical depths expressed as $\tau_{{\rm T}i}=n_{\rm e} L_i \sigma_{\rm T}$, where $\sigma_{\rm T}$ (${=\ 6.652\times10^{-25}\ {\rm cm}^2}$) is Thomson scattering cross section. Here the size of every layer is enough big so that the layers are optically thick, i.e. $L_i$ is bigger than the mean scattering length of photons. Seed photons are injected into the medium with a rate ($\dot{n}_{{\rm s}\gamma}$ {in unit of ${{\rm photons}\ {\rm cm}^{-3}\ {\rm keV}^{-1}\ {\rm s}^{-1}}$}), which is averaged in the whole corona as same as the before discussion ( Kumar \& Misra 2014, Karpouzas et al. 2019).

 Compton scattering can be considered as a diffusion process of photons
and can be described by Kompaneets equation  (Kompaneets 1957) for the low energy seed photons ($E \ll m_{\rm e}c^2$) scattered in the Comptonized medium with low energy electrons ($k T_{\rm e} \ll m_{\rm e}c^2$), where $m_{\rm e}$ is the electron mass, $c$ the light speed, $k$ the Boltzmann constant, $E$ the energy of photons. The updated form of that equation presented by Psaltis \& Lamb (1997) for a spherical corona is as follows,
 \begin{equation}
\label{eq1}
\begin{array}{cl}
t_c\frac{dn_{\gamma{}}}{dt}=&\frac{1}{{m_{\rm e}}c^2}\frac{d}{dE}\left[-4kT_{\rm e}En_{\gamma{}0}+E^2n_{\gamma{}}+kT_{\rm e}\frac{d}{dE}\left(E^2n_{\gamma{}}\right)\right]+t_c{\dot{n}}_{{\rm s}\gamma{}}-t_c\frac{c}{L}\frac{n_{\gamma{}}}{\left(1+\frac{1}{3}{\tau{}}_{\rm KN}\varepsilon\right)},
\end{array}
\end{equation}
where $t$ is time, $t_c=L/c\tau_{\rm T}$ Thomson collision timescale, $n_{\gamma}$ the photon number density per unit energy interval, $\tau_{\rm KN}=\frac{\sigma_{\rm KN}}{\sigma_{\rm T}}\tau_{\rm T}$ (i.e. $n_{\rm e} L \sigma_{\rm KN}$) the optical depth for Klein-Nishina cross section ($\sigma_{\rm KN}$) and the dimensionless parameter
 \begin{equation}
  \label{eq2}
\varepsilon=\left\{
\begin{array}{rcl}
1                                                       &      for          &   E <0.1\ m_{\rm e}c^2,\\
(1-\frac{E}{m_{\rm e}c^2}) /0.9       &      for          &  0.1\ m_{\rm e}c^2<E<m_{\rm e}c^2,\\
0                                                      &       for         &   E > m_{\rm e}c^2.\\
\end{array}
\right.
\end{equation}
 In this work, we use the units in CGS for all the variables and equations except that the unit of energy is used by keV for a concise description and calculation.

The Klein-Nishina cross section can be expressed as,
 \begin{equation}
   \label{eq3}
\begin{array}{cl}
 \sigma_{\rm KN}=&\frac{3}{4}{\sigma{}}_{\rm T}[\frac{1+x}{x^3}(\frac{2x\left(1+x\right)}{1+2x}-\ln{\left(1+2x\right))}+\frac{\ln{\left(1+2x\right)}}{2x}-\frac{1+3x}{{(1+2x)}^2}],
\end{array}
\end{equation}
where $x=\frac{E}{m_{\rm e}c^2}$. Similar to the physics described by Karpouzas et al. (2019), the probability per unit time for a photon escaping from the plasma with the depth ($L$) is $P=\frac{c}{L(1+\frac{1}{3}{\tau{}}_{KN}\varepsilon)}$ and the one from that with the depth ($L_i$) is $P_i =\frac{c}{L_i (1+\frac{1}{3}\tau_{{\rm KN }i}\varepsilon)}$. Thus the number of photons escaping from an unit volume per unit time per unit energy interval can be described by the expression $P{n_{\gamma}}$ (or $P_i{n_{\gamma i}}$).

The first expression in the right of equation (1) can be rewritten by the operator ($\hat{K}$) as,
 \begin{equation}
\label{eq4}
\begin{array}{cl}
&[\frac{1}{{m_{\rm e}}c^2}\frac{d}{dE}\left(-4kT_{\rm e}E+E^2+2 kT_{\rm e}E \frac{d}{dE}+kT_{\rm e}E^2 \frac{d}{dE}\right)] n_{\gamma{}}=\hat{K} n_{\gamma{}},
\end{array}
\end{equation}
thus equation (1) can be simplified as,
 \begin{equation}
\label{eq5}
t_c\frac{dn_{\gamma{}}}{dt}=\hat{K} n_{\gamma{}}+t_c{\dot{n}}_{{\rm s}\gamma{}}-t_c P{n_{\gamma{}}}.
\end{equation}

Comparing the whole corona with the supposed multi-layered corona, there are the same physical parameters for all plasma in the space. In addition,
there are the same number of  initial seed photons, the same energy distribution of all photons and the supposed same number of last output photons, which will be explored in Section 3. In detail, the number of the output photons per unit time per unit energy interval is,
 \begin{equation}
\label{eq6}
N=P{n_{\gamma{}}}*V=P_n{n_{\gamma{}n}}*V_{n}.
\end{equation}

In the supposed multi-layered corona, initial seed photons may be distributed in the whole corona, and also may be transported from the first layer into other layers in order, i.e. all initial seed photons are distributed only in the first layer (close to the source of the seed photons). We will discuss if the two different distributions of initial seed photons in space can lead to a little different results below.

 \subsection{Initial seed photons input into the first layer of a corona}
Firstly, all the seed photons that are input into the first layer of a corona are considered from a source (e.g. a NS). The number of the input seed photons per unit time is the same ones in the two conditions, i.e. ${\dot{n}}_{{\rm s}\gamma{}}*V={\dot{n}}_{{\rm s}\gamma{}1}*V_1$, where ${\dot{n}}_{{\rm s}\gamma{}1}$ is the injecting rate of seed photons when the initial seed photons are uniformly injected into the first layer.
The equations about the number density of  the Comptonized photons in the first layer can be written as,
 \begin{equation}
\label{eq7}
t_{\rm c}\frac{dn_{\gamma{}1}}{dt}=\hat{K} n_{\gamma{}1}+t_{\rm c}{\dot{n}}_{{\rm s}\gamma{}}\frac{V}{V_1}-t_{\rm c}P_1{n_{\gamma{}1}}.
\end{equation}
The output photons from Compton scattering in the first layer will be injected into the second layer as the seed photons. The density of the seed photons is $P_1{n_{\gamma{}1}}\frac{V_1}{V_2}$ and the equations in other layers can be expressed as,
 \begin{equation}
\label{eq8}
t_{\rm c}\frac{dn_{\gamma{}2}}{dt}=\hat{K} n_{\gamma{}2}+{t_{\rm c}P_1{n_{\gamma{}1}}}\frac{V_1}{V_2}-t_{\rm c}P_2{n_{\gamma{}2}},
\end{equation}
...
 \begin{equation}
\label{eq9}
t_{\rm c}\frac{dn_{\gamma{}i}}{dt}=\hat{K} n_{\gamma{}i}+{t_{\rm c}P_{i-1}
{n_{\gamma{}{i-1}}}}\frac{V_{i-1}}{V_i}
-{t_{\rm c}P_{i}
{n_{\gamma{}{i}}}},
\end{equation}
...
 \begin{equation}
\label{eq10}
t_{\rm c}\frac{dn_{\gamma{}n}}{dt}=\hat{K} n_{\gamma{}n0}+{t_{\rm c}P_{n-1}
{n_{\gamma{}{n-1}}}}\frac{V_{n-1}}{V_n}-{t_{\rm c}P_{n}{n_{\gamma{}{n}}}}.
\end{equation}
Combing equation (7)-(10), we obtain,
 \begin{equation}
\label{eq11}
\sum_{i=1}^n\frac{V_i}{V_n}t_{\rm c}\frac{dn_{\gamma{}i}}{dt}=\sum_{i=1}^n\frac{V_i}{V_n}
{\hat K}{n_{\gamma{}i}}+\frac{V}{V_n}t_{\rm c}{\dot{n}}_{s\gamma{}}-t_{\rm c}P_{n}{n_{\gamma{}{n}}}.
\end{equation}
Change the order of the calculation for the sum and the derivation and then divided by ${\frac{V}{V_n}}$, then equation (11) can be written as,
 \begin{equation}
\label{eq12}
t_{\rm c}\frac{d}{dt}[\sum_{i=1}^n{\frac{V_i}{V}}n_{\gamma{}i}]=
\hat{K}\left[\sum_{i=1}^n{\frac{V_i}{V}}n_{\gamma{}i}\right]
+t_{\rm c}{\dot{n}}_{s\gamma{}}-t_{\rm c}P_{n}{n_{\gamma{}{n}}}\frac{V_n}{V}.
\end{equation}

Because the distributions of photons in all the energy in the two conditions are consider to be identical, i.e. the distribution $ n_{\gamma{}}=({\sum_{i=1}^n n_{\gamma{}i}V_i})/{V}$. Substitute the expression and equation (6) into equation (12) , then equation (12) can be simplified as,
 \begin{equation}
\label{eq13}
t_{\rm c}\frac{d}{dt}n_{\gamma{}}=
\hat{K}n_{\gamma{}}
+t_{\rm c}{\dot{n}}_{s\gamma{}}-t_{\rm c}P{n_{\gamma{}}}.
\end{equation}
Due to the same expression in equation (5) and (13), it can be concluded that the solutions of the system in the two methods are equivalent when the same initial condition, energy boundary condition and space boundary condition are considered.

 According to equation (6), the relation between $n_{\gamma{}}$ and $ n_{\gamma{}n}$  can be obtained as,
 \begin{equation}
\label{eq14}
n_{\gamma{}n}=\frac{V}{V_n}\frac{L_n}{L}\frac{\left(1+\frac{1}{3}{\tau{}}_{{\rm KN}n}\varepsilon\right)}{\left(1+\frac{1}{3}{\tau{}}_{\rm KN}\varepsilon\right)}*{n_{\gamma{}}}.
\end{equation}
According to the symmetrical characteristic of equations (8), (9), (10), the number density of photons in every layer except in the first layer can be expressed as
$n_{\gamma{}i}=\frac{V}{V_i}\frac{L_i}{L}\frac{\left(1+\frac{1}{3}{\tau{}}_{{\rm KN}i}\varepsilon \right)}{\left(1+\frac{1}{3}{\tau{}}_{\rm KN}\varepsilon \right)}*{n_{\gamma{}}}$. Combining above result,  $n_{\gamma{}}=\sum_{i=1}^n{\frac{V_i}{V}}n_{\gamma{}i}$ and equation (14), the density in the first layer is expressed as,
 \begin{equation}
\label{eq15}
n_{\gamma 1}=[1-\sum_{i=2}^n \frac{L_i}{L}\frac{(1+\frac{1}{3}{\tau{}}_{{\rm KN}i}\varepsilon)}{\left(1+\frac{1}{3}{\tau{}}_{\rm KN}\varepsilon\right)}]\frac{V}{V_1}n_{\gamma}.
\end{equation}

 \subsection{Initial seed photons input into the whole corona uniformly}
Then we consider that the initial seed photons from the source are distributed into the whole corona uniformly in the beginning of the scattering. Thus the injected rate of the seed photons in every layer should include $\dot{n}_{{\rm s}\gamma}$ and the ingoing photons, which are Comptonized in the previous layer.

Similar to equation (7)-(10), the Kompaneets equations in every layer can be expressed as,
 \begin{equation}
\label{eq16}
t_{\rm c}\frac{dn_{\gamma{}1}}{dt}=\hat{K} n_{\gamma{}1}+t_{\rm c}{\dot{n}}_{{\rm s}\gamma{}}-t_{\rm c}P_1{n_{\gamma{}1}}.
\end{equation}

 \begin{equation}
\label{eq17}
t_{\rm c}\frac{dn_{\gamma{}2}}{dt}=\hat{K} n_{\gamma{}2}+t_{\rm c}{\dot{n}}_{{\rm s}\gamma{}}+{t_{\rm c}P_1{n_{\gamma{}1}}}\frac{V_1}{V_2}-t_{\rm c}P_2{n_{\gamma{}2}},
\end{equation}
...
 \begin{equation}
\label{eq18}
t_{\rm c}\frac{dn_{\gamma{}i}}{dt}=\hat{K} n_{\gamma{}i}+t_{\rm c}{\dot{n}}_{{\rm s}\gamma{}}+{t_{\rm c}P_{i-1}
{n_{\gamma{}{i-1}}}}\frac{\sum_{i=1}^i V_{i}}{V_i}
-{t_{\rm c}P_{i}
{n_{\gamma{}{i}}}},
\end{equation}
...
 \begin{equation}
\label{eq19}
t_{\rm c}\frac{dn_{\gamma{}n}}{dt}=\hat{K} n_{\gamma{}n}+t_{\rm c}{\dot{n}}_{{\rm s}\gamma{}}+{t_{\rm c}P_{n-1}
{n_{\gamma{}{n-1}}}}\frac{\sum_{i=1}^n V_{n}}{V_n}-{t_{\rm c}P_{n}{n_{\gamma{}{n}}}}.
\end{equation}
Combing equation (16)-(19), the same result in equations (13), (14) and (15) can be obtained when the same number density
of input seed photons and output photons and the same energy distribution of all photons are considered.
It means that the different distributions of initial seed photons can not lead to a big difference of number density of photons in
the two conditions.

As seen in Equation (1)-(13), (16)-(19), there is a transform invariance of the updated Kompaneets equation for the transform on layering the optically thick uniform sphere corona. According to Noether's Theorem, a transform invariance should be accompanied with a conservative quantity in physics. In this work, that the same input initial seed photons scattered in the same electrons system should produce the same output photons reflects the conservation. In another word, the conservation of the number of output photons matches the transform invariance on layering the optically thick uniform sphere corona. The transform invariance will be correct if the conservation is correct and vice versa.

However, it is difficult to match the condition because the ratio of number density of photons ($\frac{n_{\gamma n}}{n_{\gamma}}$) in equation (14) depends on energy but the parameters of the corona  in equation (6) do not depend energy. Even so, there are a few ${n_{\gamma n}(E)},\ {n_{\gamma}(E)}$ that can match the transform invariance and most of them approximately match the invariance for a LMXB. Then the prerequisite on layering the Comptonized spherical corona, i.e. the condition of the approximate conservation of output photons numbers, can be expressed as $P{n_{\gamma{}}}*V\sim P_n{n_{\gamma{}n}}*V_{n}$ according to equation (6) and (14) , which will be explored below.

\section{Compton up-scattering in a steady radiation in an uniform  spherical corona around the NS in an XB for the two conditions}
In an X-ray binaries, the spherical corona around a compact star is often discussed (Nowak et al. 1999, Barret et al. 2000, Wardzi{\'n}ski \& Zdziarski 2000).
Then we consider an example for the Compton up-scattering in a steady radiation ($\frac{dn_{\gamma{}}}{dt},\ \frac{dn_{\gamma{}i}}{dt}=0$) in the two types of a corona, i.e. a whole uniform spherical corona around a NS and two layers of the uniform spherical corona.
The NS releasing the seed photons is considered as a spherical blackbody (BB) and the injecting rate of seed photons in per unit volume of the whole corona is,
 \begin{equation}
\label{eq20}
{\dot{n}}_{s\gamma}=\frac{3R^2}{{\left(R+L\right)}^3-R^3}\left(\frac{2\pi{}}{h^3c^2}\frac{E^2}{e^{\frac{E}{kT_{b}}}-1}\right),
\end{equation}
where $T_{\rm b}$ is the temperature of BB radiation and the above expression of BB seed photons is only appropriate for the initial seed photons input into the whole corona. It is noticeable that the initial injecting rate of seed photons for the two layers of a corona is  ${\dot{n}}_{{\rm s}\gamma{}1}={\dot{n}}_{{\rm s}\gamma{}}\frac{V}{V_{1}}$, which is similar to the one in equation (7). In addition, the fully same energy natural boundary condition ($n_{\gamma},\ n_{\gamma i}=0$ at $E=2,\ 60\ {\rm keV}$ for some X-ray binaries, Zhang et al. 2017, Karpouzas et al. 2020) is considered for the whole spherical corona
and every layer of the two layers corona.

In order to test that layering the Comptonized corona is approximately feasible, firstly the distribution of the number density of photons in the corona is calculated when the same input seed photons, the same output photons and the same electrons system are considered.

As shown in Figure 1, the ratio of $n_{\gamma 1}$ obtained from the result of a multi-layered corona (i.e. equation (7)) to $n_{\gamma 1}$ obtained from the theoretical result of equation (15) depends on the energy of photons. In addition, it is also depend on the parameters of a corona, which can be seen from the curves with eight groups of parameters of a corona. The ratio is approximately equal to unit when $n_{\gamma}$ in the whole corona is substituted into equation  (15). And the same is with the ratio of $n_{\gamma 2}$ from the result of a multi-layered corona (i.e. equation (8)) to that from equation (14).
 It means that the results of number density of photons in the corona in the two supposed conditions are approximately equivalent so that the equations (equations (5), (7) \& (8) ) can be used to calculate the physical parameters on the number density of photons in every layer. The values of the ratios are closer to unit, the approximate result is better. The results from the second layer are better than that from the first layer. Thus the result from the local optical depth ratio (20:2) is a little better than that from the ratio (2:20) because the ratio of photons number density in both two layers for the optical depth ratio (20:2) are closer to unit. It can be concluded that a thick inner layer (close to the source of seed photons) will give a good match in the two condition.

In addition, we also calculated the results for two layers of a corona when the initial seed photons are uniformly distributed in whole corona (marked with diamonds in Figure 1). On the condition that the corona are considered to be multiple layers, the result from the initial seed photons distributed uniformly in the whole corona is a little worse than the one from the initial seed photons distributed only in the first layer because the ratios at low energy are different (see the points marked by diamonds and left-angles in the left panel of Figure 1). That can be understood because most initial seed photons should diffuse in the corona from the 1st to the nth layer in order. Thus the method for initial seed photons input into the first layer of a corona should be used in order to consider some physical process that happens in local part of a corona. It is noticeable that the ratios are not unit and close to unit, which means that energy distribution of photons number density in every layer in the two conditions are not identical but approximately identical.

\begin{figure*}[!htbp]
\begin{center}
\includegraphics[width=1.0\columnwidth]{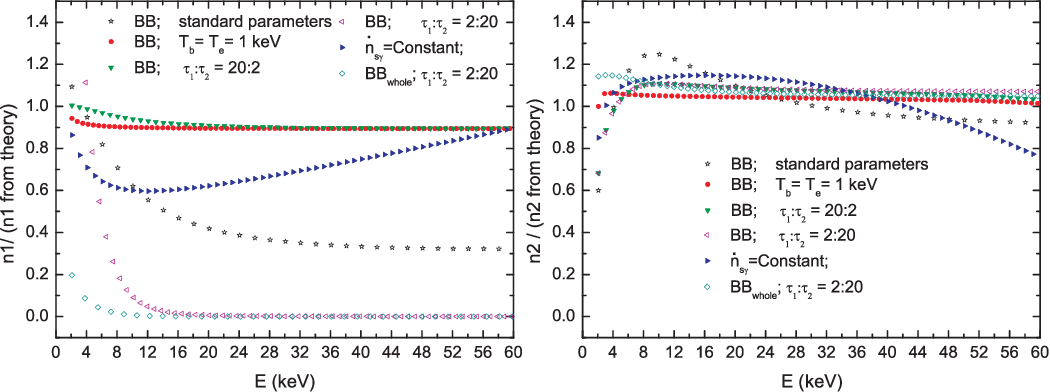}
  \caption{The relation between a ratio, which is the photon density calculated from two layers of a corona divided by the the photon density from a whole corona according to the theoretical expressions (equation14, 15), and energy of photons. The values of the ratios are closer to unit, the approximate result is better. The standard parameters of the corona are chosen as, $\tau_1=8,\ \tau_2=5,\  kT_{\rm e}=8\ {\rm keV}, kT_{\rm b}=1\ {\rm keV},\ L_{2}=2\ {\rm km}$.  Left: the first layer; Right: the second layer.}
  \label{fig1}
 \end{center}
\end{figure*}

Next the same five parameters and energy boundary conditions with the above calculation are used so as to test the basic assumption ($P{n_{\gamma{}}}*V\sim P_n{n_{\gamma{}n}}*V_{n}$ ), i.e. the approximate transform invariance.
As shown in Figure 2, the spectrums of Compton scattering observed near the surface of the corona in the two conditions are obtained when the same input photons and the same corona parameters are considered in each panel.
The output rate of photons in the two conditions are very close to each other but there are a little deviation on the right of the peaks in every panel. The ratio of the output rates for the two conditions are between 0.5 and 1.25 for all energy of photons and are $\sim 1$  for most energy region in the two layers of a corona. It means that the assumption (the same input photons scattered in the same uniform spherical corona will have the same number density of output photons) for a whole corona and two layers of a corona is approximately feasible.

\begin{figure*}[!htbp]
\begin{center}
\includegraphics[width=1.0\columnwidth]{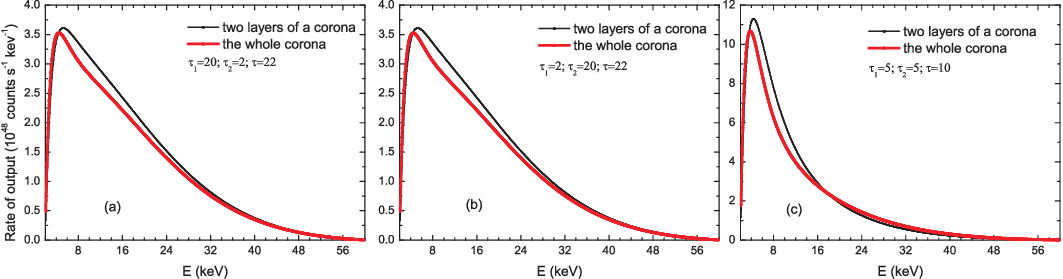}
    \caption{The spectrums of Compton scattering observed at the surface of the corona in the two conditions,  i.e. for the whole corona and two layers of a corona  when the other parameters ($kT_{\rm e}=8\ {\rm keV}$, $kT_{\rm b}=1\ {\rm keV}$, $L_{2}=2\ {\rm km}$) are choosed.  }
  \label{fig2}
 \end{center}
\end{figure*}

In order to test the effect of the five characteristic parameter of a NS-LMXB on the output photons in the two conditions, the standard value regions of the five parameters in NS-LMXBs are chosen and they are divided into 32 groups, which is shown in table 1. Then the numbers of output photons in the two structures of a corona are calculated by Monte Carlo method, in which 4000 types of combinations for the five parameters in every group are used. As shown in Figure 3, the average values of the ratios of output photons in every group, i.e. $<\frac{ P_2{n_{\gamma{}2}}*V_{2}}{P{n_{\gamma{}}}*V}>$, where $<\ >$ express the operator on average values. The average ratios for the smaller optical depth in the two layers of a corona in panel (a) are smaller than those in panel (b). It is clear that the bigger optical depth of every layer that is divided from a corona can make the approximate result ($P{n_{\gamma{}}}*V\sim P_2{n_{\gamma{}2}}*V_{2}$) better to be used in NS-LMXBs since $\frac{ P_2{n_{\gamma{}2}}*V_{2}}{P{n_{\gamma{}}}*V}\sim\ 1$. Thus the scattering from the last layer of a multi-layered corona is approximately equal to the scattering in the whole corona. In addition, the ratios of the numbers of output photons for the LMXBs with the big $kT_{\rm e}$ and $kT_{\rm b}$ are shown in panel (c).
Both temperatures of electrons and those of seed photons in groups $1\ \sim\ 8$ are smaller than those in other groups, however the two temperatures in groups $25\ \sim\ 32$ are higher than those in other groups. It can be seen that the values of those ratios in groups $25\ \sim\ 32$ are much higher than the ones in other groups except the 25th, 29th, which have smaller $\tau_1$ and $\tau_2$. In summary, the conclusion that the approximate result can be used is determined by the optical depths firstly and then by the two temperatures of electrons and seed photons.

\begin{table*}
 \centering
  \caption{The region of the five characteristic parameters that are divided into 32 groups.}
\label{t:1}
  \begin{tabular}{lccccc}
   \hline
  \hline
Ordinal number  & $\tau_1$ & $\tau_2$ & $L_2$ (km)& $kT_{\rm e}$ (keV)&$kT_{\rm b}$ (keV)\\
 \hline
1 & $1\sim5$ & $1\sim5$  &  $0.5\sim5$ & $2.5\sim 6.0$&$0.1\sim 1.2$  \\
2 & $5\sim10$ & $1\sim5$  &  $0.5\sim5$ & $2.5\sim 6.0$&$0.1\sim 1.2$  \\
3& $1\sim5$ & $5\sim10$  &  $0.5\sim5$ & $2.5\sim 6.0$&$0.1\sim 1.2$  \\
4 & $5\sim10$ & $5\sim10$  &  $0.5\sim5$ & $2.5\sim 6.0$&$0.1\sim 1.2$  \\
5 & $1\sim5$ & $1\sim5$  &  $5\sim10$ & $2.5\sim 6.0$&$0.1\sim 1.2$  \\
6 & $5\sim10$ & $1\sim5$  &  $5\sim10$ & $2.5\sim 6.0$&$0.1\sim 1.2$  \\
7& $1\sim5$ & $5\sim10$  & $5\sim10$ & $2.5\sim 6.0$&$0.1\sim 1.2$  \\
8 & $5\sim10$ & $5\sim10$  &  $5\sim10$ & $2.5\sim 6.0$&$0.1\sim 1.2$  \\
9 & $1\sim5$ & $1\sim5$  &  $0.5\sim5$ & $6.0\sim 10.0$&$0.1\sim 1.2$  \\
10 & $5\sim10$ & $1\sim5$  &  $0.5\sim5$ & $6.0\sim 10.0$&$0.1\sim 1.2$  \\
11& $1\sim5$ & $5\sim10$  &  $0.5\sim5$ &  $6.0\sim 10.0$&$0.1\sim 1.2$  \\
12 & $5\sim10$ & $5\sim10$  &  $0.5\sim5$ & $6.0\sim 10.0$&$0.1\sim 1.2$  \\
13 & $1\sim5$ & $1\sim5$  &  $5\sim10$ & $6.0\sim 10.0$&$0.1\sim 1.2$  \\
14 & $5\sim10$ & $1\sim5$  &  $5\sim10$ & $6.0\sim 10.0$&$0.1\sim 1.2$  \\
15& $1\sim5$ & $5\sim10$  &  $5\sim10$ &  $6.0\sim 10.0$&$0.1\sim 1.2$  \\
16 & $5\sim10$ & $5\sim10$  & $5\sim10$ & $6.0\sim 10.0$&$0.1\sim 1.2$  \\
17 & $1\sim5$ & $1\sim5$  &  $0.5\sim5$ & $2.5\sim 6.0$&$1.2\sim 2.5$  \\
18 & $5\sim10$ & $1\sim5$  &  $0.5\sim5$ & $2.5\sim 6.0$&$1.2\sim 2.5$  \\
19 & $1\sim5$ & $5\sim10$  &  $0.5\sim5$ & $2.5\sim 6.0$&$1.2\sim 2.5$ \\
20 & $5\sim10$ & $5\sim10$  &  $0.5\sim5$ & $2.5\sim 6.0$&$1.2\sim 2.5$  \\
21 & $1\sim5$ & $1\sim5$  &  $5\sim10$ & $2.5\sim 6.0$&$1.2\sim 2.5$  \\
22 & $5\sim10$ & $1\sim5$  &  $5\sim10$ & $2.5\sim 6.0$&$1.2\sim 2.5$  \\
23 & $1\sim5$ & $5\sim10$  &  $5\sim10$ & $2.5\sim 6.0$&$1.2\sim 2.5$ \\
24 & $5\sim10$ & $5\sim10$  &  $5\sim10$ & $2.5\sim 6.0$&$1.2\sim 2.5$ \\
25 & $1\sim5$ & $1\sim5$  &  $0.5\sim5$ & $6.0\sim 10.0$&$1.2\sim 2.5$  \\
26 & $5\sim10$ & $1\sim5$  &  $0.5\sim5$ & $6.0\sim 10.0$&$1.2\sim 2.5$  \\
27 & $1\sim5$ & $5\sim10$  &  $0.5\sim5$ & $6.0\sim 10.0$&$1.2\sim 2.5$ \\
28 & $5\sim10$ & $5\sim10$  &  $0.5\sim5$ & $6.0\sim 10.0$&$1.2\sim 2.5$\\
29 & $1\sim5$ & $1\sim5$  &  $5\sim10$ & $6.0\sim 10.0$&$1.2\sim 2.5$  \\
30 & $5\sim10$ & $1\sim5$  &  $5\sim10$ &$6.0\sim 10.0$&$1.2\sim 2.5$  \\
31 & $1\sim5$ & $5\sim10$  &  $5\sim10$ & $6.0\sim 10.0$&$1.2\sim 2.5$ \\
32 & $5\sim10$ & $5\sim10$  &  $5\sim10$ & $6.0\sim 10.0$&$1.2\sim 2.5$ \\
\hline
\end{tabular}
\end{table*}

\begin{figure*}[!htbp]
\begin{center}
\includegraphics[width=1.0\columnwidth]{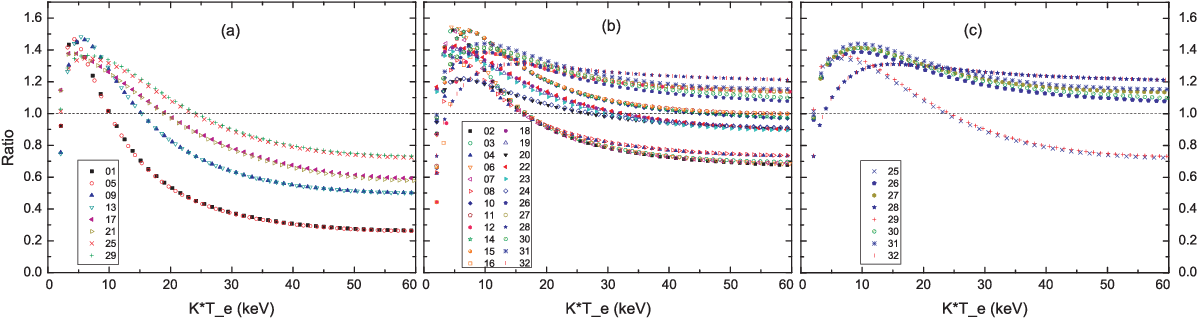}
    \caption{The average ratio of the numbers of output photons at the surface of the corona of the NS in a LMXB in the special region of the five parameters in the two conditions. Left: $\tau_1,\ \tau_2\ \in\ (1,\ 5)$;  Middle: $\tau_1$ or $\tau_2\ \in$ $(5,\ 10)$,  $\tau_1$ and $\tau_2\ \in$ $(5,\ 10)$ for the region of all parameters in this work; Right: $kT_{\rm e}\ \in\ (6,\ 10)\ {\rm keV}$, $kT_{\rm b}\ \in\ (1.2,\ 2.5)\ {\rm keV}$.}
  \label{fig2}
 \end{center}
\end{figure*}

As shown in the left panel of Figure 4, the ratios in every curves in Figure 3 are re-averaged for all energy, i.e. from $2\ {\rm keV}$ and $60\ {\rm keV}$. Except the 1th, 5th group with smaller $\tau_1$, $\tau_2$, $kT_{\rm e}$ and $kT_{\rm b}$, the re-averaged ratios in other groups are between 0.75 $\sim$ 1.25. With the increasing of $kT_{\rm e}$ and $kT_{\rm b}$, the  re-averaged ratios are discriminated obviously for the first 8 groups and the last 8 groups, which has different $kT_{\rm e}$ and $kT_{\rm b}$. In a word, the layers with big optical depth, a small electrons' temperature in a corona and a source with a small seed photons' temperature can match the approximate result ($P{n_{\gamma{}}}*V\sim P_2{n_{\gamma{}2}}*V_{2}$) well, of course the approximate result can also be used for the layers with big optical depth in a corona in other groups. In the right panel, the average values of the ratios for the total output photons and the total output energy from the Comtonized corona in the two conditions, i.e. $<\frac{\int_2^{60}P_2 n_{\gamma 2} dE*V_2}{\int_2^{60}P n_{\gamma}dE*V}>$ and $<\frac{\int_2^{60}P_2 n_{\gamma 2} EdE*V}{\int_2^{60}P n_{\gamma}EdE*V}>$, are close to each other and $\sim\ 1.8$, which can be used as an adjusted factor for the radiation flux and energy. According to the Monte Carlo calculation, the same input initial seed photons scattered in the same electrons system can approximately produce the same numbers of output photons for Compton scattering, i.e. there is an approximate transform invariance of layering the optically thick uniform sphere Comptonization corona around the neutron star in an X-ray binary.

\begin{figure*}[!htbp]
\begin{center}
\includegraphics[width=1.0\columnwidth]{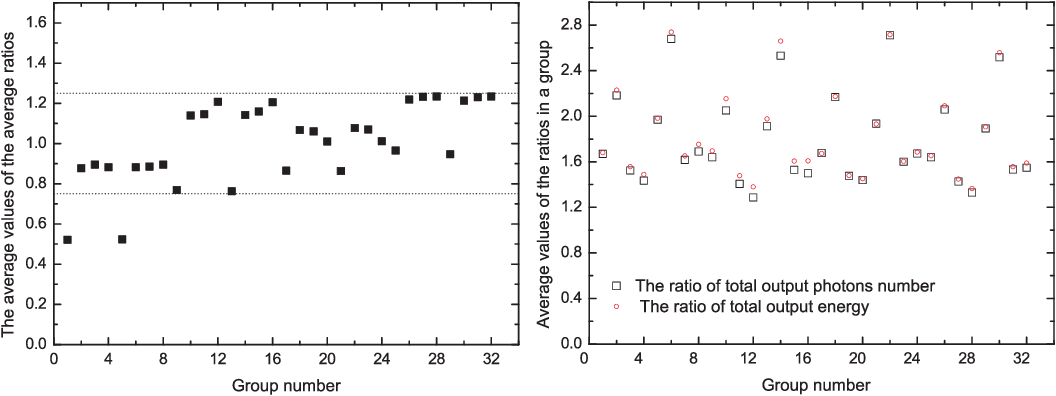}
  \caption{Left: The re-averaged ratios of the numbers of output photons at the surface of the corona of the NS in a LMXB in the special region of the five parameters in the two conditions for all energy;
  right: the average values of the ratios for the total output photons and the total output energy from the Comtonized corona in the two conditions.}
  \label{fig2}
 \end{center}
\end{figure*}

\section{DISCUSSION and CONCLUSION}
As seen in Section 3,  the number density of output photons in the two conditions are close to each other while the densities are not fully identical, i.e.the basic assumption is approximate and thus the ratios in Figure 1 are close to unit but not fully equal to unit. There may be two factors that lead to the difference of the spectrum in the two conditions.

First the last escape term in equation (1) is an approximation to the spatial diffusion part of the original Kompaneets equation including space coordinates. It is not valid if the average scattering length is of the same order as or bigger than the size of the system thus the corona can not be divided into infinite layers. In another word, every layer should be optically thick in order to match the condition that the size of every layer is much bigger than the mean scattering length, which matches our Monte carlo simulations. In addition, the more layers of the corona divided by us, and then maybe the more deviation of the number density of output photons between in the whole corona and in the multi-layered corona because of the accumulation effect of the multiple approximation. In brief, the more times that the approximate escape term is used, maybe the more deviation will be produced. In a word,  it is noticeable that the result in this work can only be applied to the optically thick system.

Second it was discussed in many works that the space distribution of initial seed photons were uniform but it may not match the transporting process of initial seed photons, i.e. it should be an ideal assumption. The nonuniform distribution of initial seed photons may lead to different scattering times for the seed photons at the different place of the corona. Thus there will be a different distributions of number densities of photons in every layer in the two conditions, however they are approximately similar in the corona around the NS in a LMXB (i.e. $ n_{\gamma{}}(E) V\ \sim\ {\sum_{i=1}^n n_{\gamma{}i}(E) V_i}$).
Thus the solutions of the equations from combining several Kompaneets equations by using the space distribution of nonuniform seed photons are not fully identical to the solutions of the source equations with the uniform seed photons but they approximately equal to the solutions of these equations in every optical thick layers of a corona. The conclusion that the ratio of number densities of photons in Figure (1) is not unit expresses the difference, however the result about that the ratio of number densities of photons is close to unit means that this approximate method is effective.

In addition, the spectrum shape of the output photons for Comptonization is complicated. According to the parameters chosen for the up-scattering in Section 3, the scattering process is an unsaturated Comptonization for the medium optical depth (${\tau\ >\ 1}$ and $ \sim$ a few). Although energy change of a photon in a single scattering is small, the black body ``seed'' input spectrum processed by many Compton scattering processes in Figure 2 will be changed. In this kind of Comptonization, the scattering between
the photons and electrons can not change the total number of the photons, but can change the mean energy.
Thus the peak will be shifted after several scatterings. Then the hard X-ray branch in the spectrum for the unsaturated Comptonization should descend with a power law shape.

In this work we obtained that Compton scattering in the optically thick uniform spherical corona of an X-ray binary for the low energy seed photons scattered in low energy electrons is approximately equivalent in the two conditions, i.e. scattering in every layer of a multi-layered corona in order and scattering in the whole corona that can be divided into several layers.
The number density of photons in every layer can be calculated from two methods for fewer layers, the direct calculation from the number density of photons in every layer of a corona and the indirect calculation from the number density of photons in the whole corona.When the multiple layers of a multi-layered corona are considered, the result for that the initial seed photons are distributed in the first layer in the beginning of the scattering is better than the one for the seed photons distributed in the whole corona. In addition, the conclusion that the same input initial seed photons scattered in the same electrons system can approximately produce the same numbers of output photons for Compton scattering indicates that there is an approximate transform invariance of layering the optically thick uniform sphere Comptonized corona around the neutron star in an XB.


Compton scattering for a corona divided into several layers may be used in many celestial bodies. There are some conditions for plasma that need to be divided into several layers such as the sandwich-like thick accretion disc around a compact star, the multiple layers of a corona around a NS. The Compton scattering for a corona to be divided into several layers may be applied into an X-ray burst for the color corrected factor (Suleimanov et al. 2011). And it can be used to discuss the oscillation process in different layers of the corona around a NS.

\section*{acknowledgements}
This work is supported by the National Natural Science Foundation of China under grant Nos. 12063001, 11563003 and by the program of China Scholarships Council (No. 201907565014).

\section*{Data availability}
The data underlying this article are available in the article and in its online supplementary material.

\clearpage

\label{lastpage}

\end{document}